          [1999/12/02 1.01 (PWD)]
\documentclass[a4paper,twocolumn]{esapub} 

\usepackage{epsfig} 		
\usepackage{natbib}

\title{COMPTEL Observations of the Blazar PKS 1622-297
       during a Gamma-Ray High State in 1995}
\author[1,6]{ S.~Zhang}
\author[1]{W.~Collmar}
\author[1]{V.~Sch\"onfelder}
\author[2]{H.~Bloemen}
\author[2]{W.~Hermsen} 
\author[3]{\\J.~Ryan}
\author[4]{K.~Bennett}
\author[4]{O.R.~Williams}
\author[5]{O.~Reimer}
\affil[1]{Max-Planck-Institut f\"ur extraterrestrische Physik, Garching,
Germany}
\affil[2]{Space Research Organization Netherlands, Utrecht, The Netherlands}
\affil[3]{Space Science Center, University of New Hampshire, Durham, USA}
\affil[4]{Astrophysics Division, ESTEC, Noordwijk, The Netherlands}
\affil[5]{NASA/Goddard Space Flight Center, Greenbelt, MD 20771, USA
}
\affil[6]{High Energy Astrophysics Lab, IHEP, P.O.Box 918-3, Beijing, China
}

\begin{document}


\maketitle

\begin{abstract}
PKS~1622-297 was detected as a source of $\gamma$-rays by the EGRET
 experiment at energies above 100 MeV
during a $\gamma$-ray outburst June and July 1995.
We analyzed the COMPTEL data (0.75-30 MeV) of this time period
to investigate the behaviour of PKS 1622-297 at lower $\gamma$-ray energies.
The blazar is significantly (5.7$\sigma$) detected by COMPTEL at energies above 10 MeV. Below 10~MeV the source is only marginally (3-10~MeV band)
or not (below 3~MeV) detected. The summed MeV spectrum shows a 'hard'
($\alpha_{ph}$ $<$2, $\sim$$E^{-\alpha}$) shape and, if combined with 
the simultaneous EGRET spectrum, a spectral break at MeV energies
is indicated. 
 We present the COMPTEL results 
(light curves, spectra) and compare them to results derived in neighbouring energy bands, in particular to the EGRET one.
\end{abstract}

\section{Introduction}

The EGRET experiment aboard CGRO has identified more than
60 blazar-type AGN emtting at high-energy ($>$100~MeV) $\gamma$-rays
(Hartman et al. 1999). Most of them show a time-variable $\gamma$-ray
flux and a few even strong $\gamma$-ray flares. 
One such example is PKS~1622-297, which was detected 
by EGRET during a $\gamma$-ray high state in June and July 1995,
and on top of that showed a strong $\gamma$-ray flare lasting for 
a few days (Mattox et al. 1997). 
After the $\gamma$-ray activity of PKS~1622-297 was recognized by EGRET 
a target-of-opportunity (ToO) observation for CGRO was scheduled which 
led to the detection of PKS~1622-297 also at hard X- to soft $\gamma$-rays
by the OSSE experiment (Kurfess et al. 1995). 

The COMPTEL experiment (Sch\"onfelder et al. 1993), measuring in
the MeV band between 0.75 and 30~MeV, has detected 9 blazars
(e.g., Collmar et al. 1999). Among them is PKS~1622-297
which was detected during this $\gamma$-ray high state period in 1995. 
Preliminary MeV results have been reported by
Collmar et al. (1997).
In this paper we briefly present first results of a detailed analysis
of the COMPTEL data of this period, and compare them to results in
neighbouring energy bands, in particular to the EGRET band.  
A more detailed presentation of the results of the analysis will be 
given in Zhang et al. (2001). 

\begin{table}[h]
  \begin{center}
    \caption{~COMPTEL observations of PKS 1622-297 during its 4-week
$\gamma$-ray high state in 1995. The CGRO VPs and their time periods
 are given.
The offset angle refers to the angle between the pointing direction of the instrument and the source position.}\vspace{1em}
    \renewcommand{\arraystretch}{1.2}
\begin{tabular}{ccc}
\hline
VP & Date & Offset angle\\ \hline 
421.0 & 06/06/95-13/06/95& 14.5$^{\circ}$  \\ 
422.0 & 13/06/95-20/06/95  & 15.2$^{\circ}$ \\ 
423.0 & 20/06/95-30/06/95  & 19.2$^{\circ}$ \\ 
423.5 & 30/06/95-10/07/95  & 3.0$^{\circ}$ \\ 
\hline
\end{tabular}
    \label{tab:1}
  \end{center}
\end{table}


\section{OBSERVATION AND DATA ANALYSIS}

PKS~1622-297 was observed during a four-week observations towards
the Galactic Center region from June 6 to July 10, 1995 covering the 
CGRO viewing periods (VPs) 421 to 423.5. For more details on the 
observations see Tab.~1. 

The analyses were carried out using the standard COMPTEL 
maximum-likelihood analysis procedure (e.g., de Boer et al. 1992)
including a filtering technique 
for background generation (e.g., Bloemen et al. 1994) to 
derive detection significances, fluxes and flux errors 
of the $\gamma$-ray source in the four standard COMPTEL energy
bands. Diffuse emission models have been included in the analysis 
to subtract off the diffuse galactic $\gamma$-ray emission.  

\begin{figure}[tb]
\centering
\psfig{figure=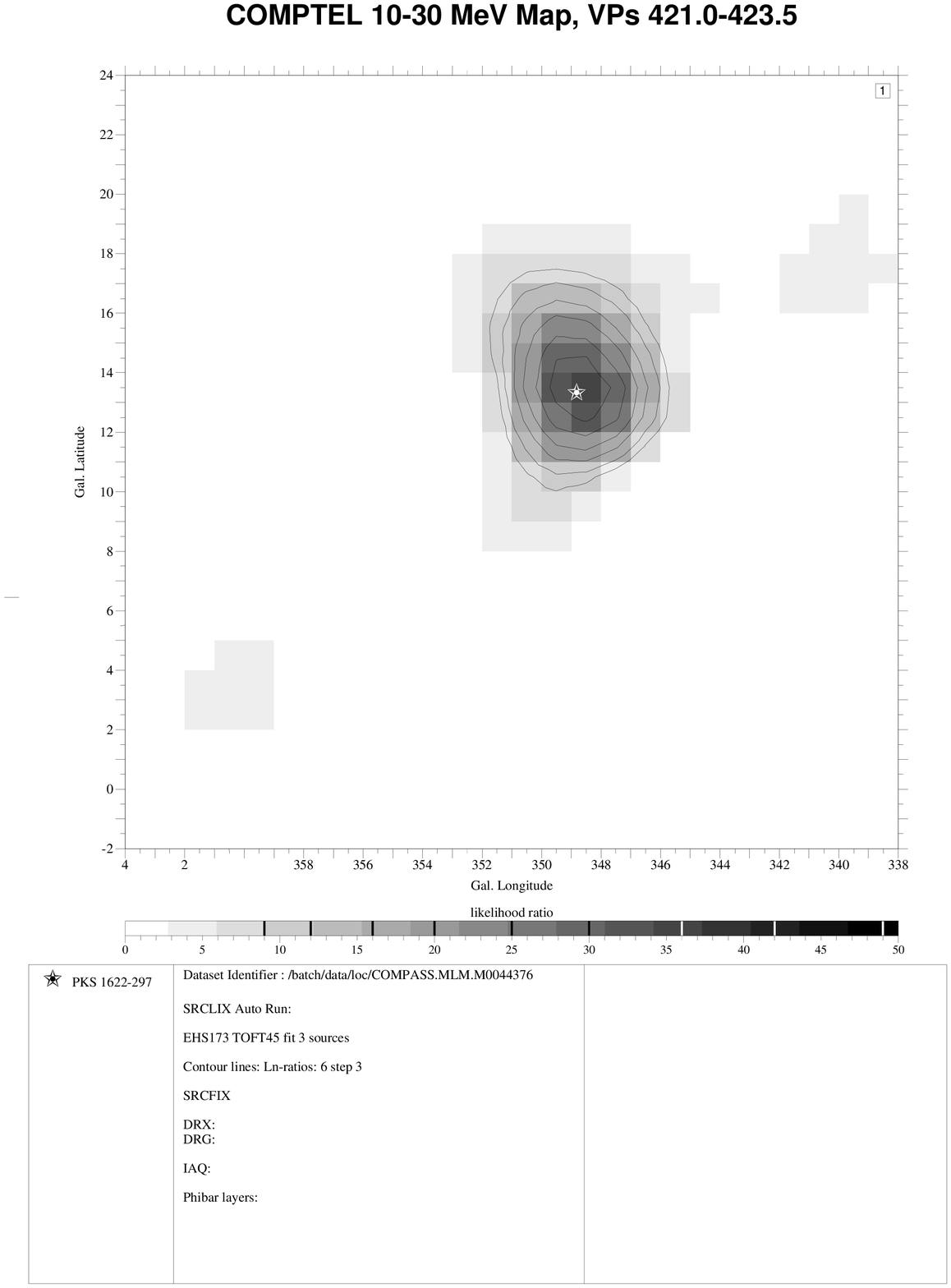,height=8cm,bbllx=2.0cm,bblly=7.5cm,bburx=19.5cm,bbury=25.9cm,clip=}
\caption{COMPTEL 10-30 MeV map for the 4-week $\gamma$-ray 
high state of PKS 1622-297 ($\star$) 
in 1995. The contour lines start at a detection significance level
 of 3$\sigma$ 
with steps of 0.5$\sigma$.\label{fig:single}}
\end{figure}

\section{Results}
\subsection{Detections}

In the sum of the 4 weeks, PKS~1622-297 is detected with a significance 
of $\sim$5.7$\sigma$ in the upper -- 10 - 30~MeV -- COMPTEL band (Fig.~1).
Below 10 MeV the source becomes weak. In the 3 - 10~MeV band only a hint is 
obtained, and below 3~MeV the source is not detected anymore.
Above 10~MeV there is evidence for the source in all 4 individual VPs
showing that PKS~1622-297 was emitting MeV $\gamma$-rays 
for the whole period. However, the detection significances
with about 2 to 3$\sigma$ each, become weak.  
The relevant fluxes and upper limits are given in Tab.~2.  

\begin{table}[bh]
  \begin{center}
    \caption{~Fluxes and upper limits for PKS 1622-297 for flare state
 (VPs 421.0-423.5) in units of 10$^{-5}$ ph cm$^{-2}$ s$^{-1}$.
 The energy bands are given in MeV. The errors are 
1$\sigma$ and the upper limits are 2$\sigma$. An upper limit is given when 
the significance of an individual flux value is less than 1$\sigma$.}\vspace{1em}
    \renewcommand{\arraystretch}{1.2}
    \label{tab:2}
\begin{tabular}{ccccc}
\hline
  Period (VP) & 0.75-1 & 1-3 & 3-10 & 10-30   \\
\hline
421.0 & $<$28.3 & $<$25.2 & $<$10.7 & 3.4$\pm$1.7 \\
422.0 & 18.3$\pm$13.6 & $<$28.0 & $<$8.9 & 3.2$\pm$1.5      \\
423.0 & 18.1$\pm$12.5 & $<$23.9 & $<$10.0 & 5.1$\pm$1.5  \\
423.5 & $<$21.5 & $<$20.8 & $<$9.7 & 1.8$\pm$1.0 \\
421.0-423.5 & $<$17.4 & $<$11.2 & 2.0$\pm$1.9 & 3.5$\pm$0.7 \\
\hline
\end{tabular}  \end{center}
\end{table}

\begin{figure}[th]
\centering
\psfig{figure=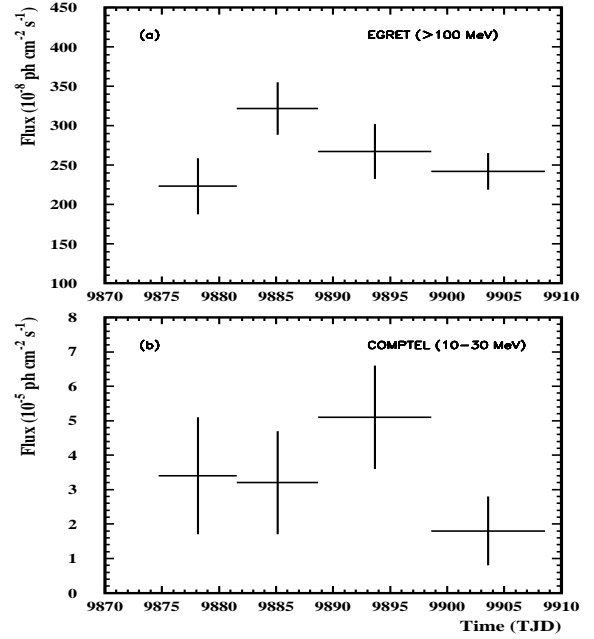,height=9.0cm,width=8.0cm,clip=}
\caption{Light curves for PKS 1622-297 as observed by EGRET
 at energies $>$ 100 MeV (a) and by COMPTEL in the energy range 10-30 MeV 
(b) during the 4 individual VPs of the high state in 1995. The EGRET
 data are
from Hartman et al. (1999). The error bars are 1 $\sigma$.
\label{fig:2}}
\end{figure}

\begin{figure}[tbh]
\centering
\epsfig{figure=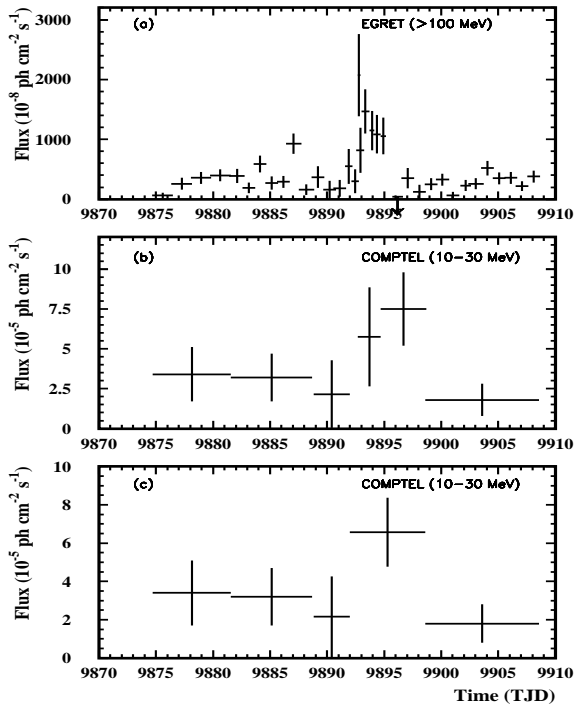,height=10.0cm,width=8.0cm,clip=}
\caption{High time resolution EGRET light curve for PKS 1622-297 (a) and compared 
to COMPTEL light curves (b, c), where the major flaring period is better 
resolved in time.  Fig. 3c is the same as Fig. 3b with the two high fluxes combined to show the enhanced MeV emission around the time of EGRET major flare. The EGRET data are from Mattox et al. (1997).\label{fig:3}}
\end{figure}

\begin{figure}[tbh]
\centering
\epsfig{figure=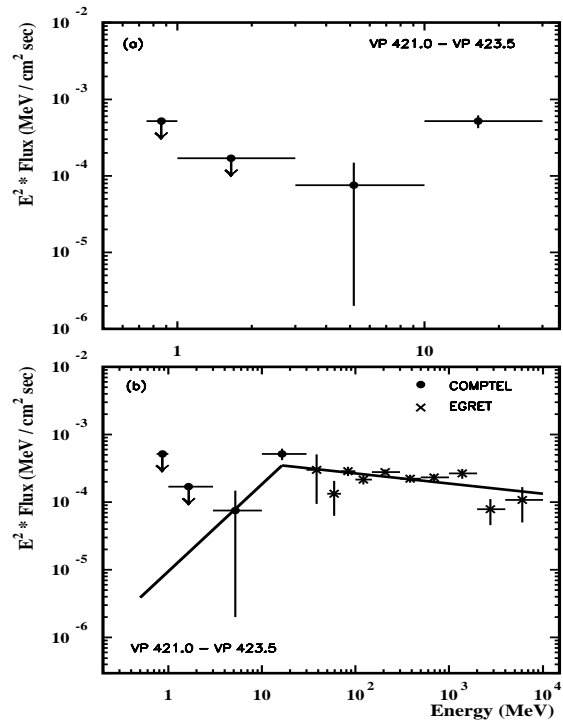,height=10.0cm,width=8.0cm,clip=}
\caption{The average COMPTEL spectrum (a) and the combined 
COMPTEL/EGRET spectrum (b) together with the best-fitting broken 
power-law model for the 4-week $\gamma$-ray high-state
period are shown. The evidence for the spectral break around
$\sim$20~MeV is obvious.\label{fig:4}}
\end{figure}

\subsection{Time Variability}
The EGRET ($>$100~MeV; from Hartman et al. 1999)
 and COMPTEL 10-30~MeV light curve of the 
4 individual CGRO VPs show neither for 
COMPTEL nor for EGRET strong evidence for time variability (Fig.~2).
However, on shorter time scales ($\sim$1~day)
Mattox et al. (1997) reports strong time variability 
including an major flare event on top 
of the $\gamma$-ray high state, which occurred within VP~423 
and lasted for about 2 days (Fig.~3).
We searched the COMPTEL data for a possible simultaneous
MeV flare, by subdividing VP 423 according to the EGRET light curve
in 3 parts: a pre-flare (TJD 9888.8 - 9892.7), on-flare
 (TJD 9892.7 - 9894.7), and post-flare part (TJD 9894.7 - 9898.6).
The fluxes for the on-flare 
and post-flare part are higher than the other 10-30~MeV fluxes 
indicating the presence of the flare also at MeV energies (Fig.~3).
We note, however, that the highest COMPTEL flux is found after the flare
 in the 
EGRET band had already vanished. This indicates a possible time delay of a few 
days between the two bands in the sense that the higher energies come 
first. 

\subsection{Energy Spectra}
We have added the COMPTEL data of the 4-week high-state period to 
generate an average MeV spectrum. Due to the non-detections below 3 MeV
only 2 spectral points are derived, which, nevertheless, indicate 
a hard ($\alpha_{ph}$ $<$2, $\sim$$E^{-\alpha}$) spectrum of 
PKS~1622-297 at MeV energies (Fig.~4). 
We also generated the simultaneous EGRET spectrum of PKS~1622-297, 
which is between 30~MeV and 10~GeV sufficiently fitted by a 
power-law shape with a photon index of $\sim$2.2. If the simultaneous 
EGRET and COMPTEL spectra are combined (Fig.~4), 
a spectral break at MeV energies becomes obvious. 
A fit of this simultaneous EGRET and COMPTEL spectrum with a broken 
power-law model indicates the break energy to be around 20~MeV. 
Due to the marginal COMPTEL detections below 10~MeV, the amount of
the spectral break, i.e. the change in photon index from the higher
to the lower energies, is not well determined. Its lower limit, however, 
should be larger than 0.6.    
 
\begin{figure}[tbh]
\centering
\epsfig{figure=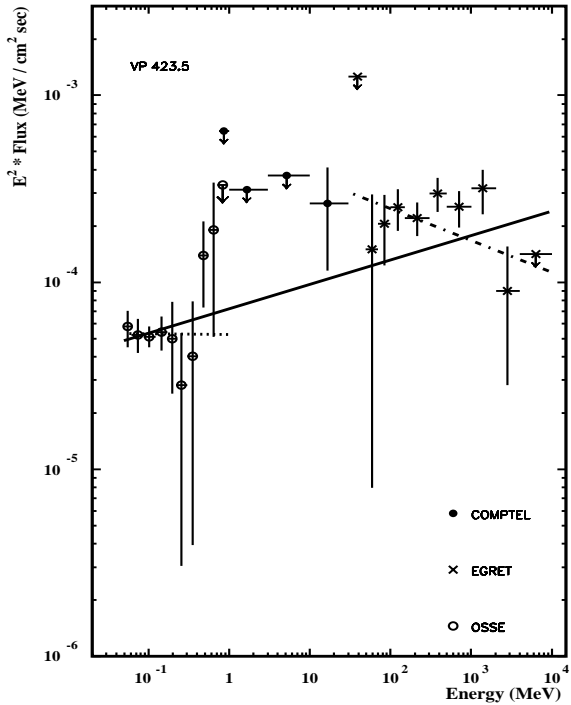,height=10.0cm,width=8.0cm,clip=}
\caption{The combined simultaneous OSSE/COMPTEL/EGRET spectrum for 
the ToO period (VP 423.5).
The EGRET and OSSE data and the relevant fit lines are from Mattox et al. (1997). The dotted line symbolizes the OSSE fit, the dashed-dotted line 
the EGRET fit, and the solid line the combined EGRET/OSSE fit.\label{fig:5}}
\end{figure}

After the $\gamma$-ray activity of PKS~1622-297 had been recognized
by EGRET, a ToO observation (VP~423.5) was carried out, where
OSSE participated. This provided the possibility to measure the 
spectrum down to 0.05~MeV. 
Mattox et al. (1997) reported that the OSSE and
EGRET data itself can be represented by power-law shapes with similar
photon indices of 2.0$\pm$0.2 and 2.2$\pm$0.1, respectively (Fig.~5).
However, fitting a simple power-law shape to the 
combined OSSE/EGRET data, they obtained a photon spectral index of
1.87$\pm$0.02.
We have analysed the COMPTEL data of VP~423.5 and have added 
the spectral results 
to this combined OSSE-EGRET spectrum (Fig.~5).
Although these COMPTEL data are not really 
constraining, they -- at first glance -- support a more complex spectral 
shape at $\gamma$-ray energies than the simple power-law model.

\section{Discussion}

In general, PKS~1622-297 shows a similar behaviour as is observed 
in other such sources, like PKS~0528+134 and 3C~279 for example. 
During flaring events observed by EGRET, PKS~1622-297 is detected
 -- predominantly --
at the upper COMPTEL energies, resulting in a 'hard' MeV spectrum
 and showing  a spectral break in the 
combined COMPTEL-EGRET spectrum. These facts indicate that the flare  
is more efficient at higher $\gamma$-ray energies. 
Such a behaviour can generally be explained in a multi-component 
$\gamma$-ray emission scenario, consisting of synchrotron-self Compton and 
external Comptonization components. Because they have different 
dependencies on the bulk Lorentz factor, a change in bulk Lorentz factor 
could result in increased emission at mainly the higher-energy 
$\gamma$-rays  (e.g., B\"ottcher $\&$ Collmar 1998).

On top of the already high $\gamma$-ray level, PKS~1622-297 showed 
a major flare which lasted for a few days. COMPTEL measures enhanced 
$\gamma$-ray emission during this period. This suggests that 
the flare is also visible at MeV energies, however with reduced amplitude.
An intriguing feature of this flare is the possible time delay of a few days between
the $\gamma$-rays above 100~MeV and the 10-30 MeV band, indicating that 
the opacity of the $\gamma$-ray emitting region might be different 
for the different energy bands. Such a delay has not yet been observed in 
another blazar. A measurement of this delay, together with the ratio
of the flare amplitudes in the different bands, has the potential to 
improve our insights on the $\gamma$-ray emitting region.

\section{Summary}

During a 4-week $\gamma$-ray acitivity period observed at energies above 
100~MeV by EGRET, PKS~1622-297 is also significantly detected at
lower-energy $\gamma$-rays  ($>$10~MeV) by COMPTEL. This detection together
with the marginal detection and non-detection at lower COMPTEL
energies indicates a 'hard' ($\alpha_{ph}$ $<$2, $\sim$$E^{-\alpha}$)
MeV spectrum. The combined COMPTEL-EGRET spectrum requires a 
spectral break around $\sim$20~MeV.  

On top of this $\gamma$-ray high state, EGRET observed a 
major flare which lasted for about 2~days. COMPTEL observes 
enhanced MeV emission (10 - 30 MeV band) during that time. However, the 
COMPTEL flux is still high when the EGRET one is already down, 
indicating a possible time delay of a few days between the two bands. 

\section*{Acknowledgements}
This research was supported by the German government through DLR grant 50 QV 9096 8, by NASA under contract NAS5-26645, and by the Netherlands Organisation for Scientific Research NWO.

\end{document}